# Innovation as a Nonlinear Process, the Scientometric Perspective, and the Specification of an "Innovation Opportunities Explorer"

*Technology Analysis and Strategic Management* (forthcoming in 2013),

Loet Leydesdorff,* Daniele Rotolo,** & Wouter de Nooy***

*Amsterdam School of Communication Research (ASCoR), University of Amsterdam, Kloveniersburgwal 48, 1012 CX Amsterdam, The Netherlands; loet@leydesdorff.net.*

**SPRU (Science and Technology Policy Research), University of Sussex, Brighton, East Sussex, United Kingdom; d.rotolo@sussex.ac.uk.*

***Amsterdam School of Communication Research (ASCoR), University of Amsterdam, Amsterdam, The Netherlands; w.denooy@uva.nl.*

**Loet LEYDESDORFF** is Professor at the Amsterdam School of Communications Research (ASCoR) of the University of Amsterdam, Visiting Professor of the Institute of Scientific and Technical Information of China (ISTIC) in Beijing, and Honorary Fellow of the Science and Technology Policy Research Unit (SPRU) of the University of Sussex. With Henry Etzkowitz, he initiated a series of conferences, workshops, and special issues about the Triple Helix of University-Industry-Government Relations.

**Daniele ROTOLO** is research fellow at SPRU–Science and Technology Policy Research (University of Sussex). He was visiting researcher at University College London (UCL) and Stern Business School–New York University. Daniele is currently newsletter editor of the Technology and Innovation Management (TIM) division of the Academy of Management. His research interests lie in the social capital theory and social network analysis. More precisely, adopting these theories as lens he is investigating the social dynamics characterizing the science and technology interaction and the emerging technologies.

**Wouter DE NOOY** is Associate Professor at the Amsterdam School of Communication Research (ASCoR) of the University of Amsterdam. He specializes in social network analysis, dynamic statistical network models in particular. He is one of the authors of *Exploratory Social Network Analysis with Pajek* (Cambridge University Press).

# Innovation as a Nonlinear Process, the Scientometric Perspective, and the Specification of an Innovation Opportunities Explorer

The process of innovation follows non-linear patterns across the domains of science, technology, and the economy. Novel bibliometric mapping techniques can be used to investigate and represent distinctive, but complementary perspectives on the innovation process (e.g., "demand" and "supply") as well as the interactions among these perspectives. The perspectives can be represented as "continents" of data related to varying extents over time. For example, the different branches of Medical Subject Headings (MeSH) in the Medline database provide sources of such perspectives (e.g., "Diseases" *versus* "Drugs and Chemicals"). The multiple-perspective approach enables us to reconstruct facets of the dynamics of innovation, in terms of selection mechanisms shaping localizable trajectories and/or resulting in more globalized regimes. By expanding the data with patents and scholarly publications, we demonstrate the use of this multi-perspective approach in the case of RNA Interference (RNAi). The possibility to develop an "Innovation Opportunities Explorer" is specified.

**Introduction**

A scientometric perspective on innovations is difficult to obtain because innovations almost by definition occur across scientific, technological, and economic domains that have been archived using different databases and classifications, and hence from different, but possibly interacting perspectives. Whereas scientometrics has focused on output indicators of the science and technology system such as publications, citations, and patents, economists can consider patents and other knowledge carriers as input to "total factor productivity" (TFP; Solow, 1957; e.g., Coe *et al.*, 2009). As Grilliches (1994, at p. 14) pointed out in his Presidential Address to the American Economic Association: "Our current statistical structure is badly split, there is no central direction, and the funding is heavily politicized." In order to solve the ensuing "computer paradox" that has made the measurement problems worse despite the abundance of data, Grilliches placed his hope at the time on the efforts of Jaffe *et al.* (1993), Trajtenberg (1990), and others to use patent citations as a tool to measure the dynamics of knowledge and innovation in the economy (cf. Grilliches, 1984).

In their *magnum opus* entitled "Patents, Citations, and Innovations: A Window on the Knowledge Economy," Jaffe & Traitenberg (2002) elaborated on their analysis of the database of the U.S. Patent and Trade Organization (USPTO) using almost three million US patents granted between January 1963 and December 1999, and more than 16 million citations of these patents between 1975 and 1999. Despite the ambitious title of the project, however, patent-based measures do *not* capture innovations, while patents are indicators of inventions. The primary function of patents is legal protection against litigation in court; examiners add references to the knowledge claims for the purpose of showing novelty in relation to prior art (Alcácer *et al.*, 2009; Criscuolo & Verspagen, 2008), but not on the basis of (future) market potential.

In parallel to these efforts, the Organization of Economic Co-operation and Development (OECD) in Paris organized a workshop in 1994 entitled "Employment and Growth in the Knowledge-Based Economy," also in response to the critique from governments of member states and notably the European Commission that the organization had hitherto used the framework of "*national* systems of innovation" (Freeman, 1987; Lundvall, 1988; Nelson, 1993). Since the demise of the Soviet Union (1991) and the opening of China, *globalization* across national boundaries had become a major factor driving economic change. At this workshop, Abramowitz and David (1996, p. 35) suggested that *codified* knowledge should be made central to the analysis of a knowledge-based economy (cf. Dasgupta & David, 1994). Had "a new fusion between science and innovation" historically become possible in a post-industrial society (Bell, 1968, at p. 182)?

In our opinion, knowledge-based coordination tends to transform political economies. Whereas political economies are based on two coordination mechanisms—economic market mechanisms and political regulation—and are nationally organized and equilibrium-oriented (Aoki, 2001), knowledge-based economies are based on three interacting coordination mechanisms: wealth generation in the market, institutional control by political agency, and novelty production in science and technology (Leydesdorff, 2006, 2010a). Three sub-dynamics may lead to meta-stabilization, hyper-stabilization, and also globalization at a next-order systems level.

A neo-evolutionary version of the Triple Helix model can thus be developed and juxtaposed with a neo-institutional version of university-industry-government relations (Leydesdorff & Zawdie, 2010). In the neo-institutional model, the network arrangements can be analyzed in terms of their efficiency and efficacy in institutional learning (Etzkowitz, 2008), but the neo-evolutionary model focuses on the specification of selection environments and their interactions. Selections operate asymmetrically among domains, and selection mechanisms (at the structural level) can be expected to shape recursively nonlinear trajectories and regimes (Dosi, 1982).

After the workshop in 1994, the OECD provided funding for a new program to develop indicators of 'the knowledge-based economy' (David & Foray, 1995; OECD, 1996). This led to the regular publication of the *Science, Technology, and Industry Scoreboards*,[1] and a periodic summary of progress at the ministerial level (cf. Foray, 2004; David & Foray, 2002).[2] Godin (2006, at p. 24) noted that the metaphor of a 'knowledge-based economy' has functioned, in this context, mainly as a label for reorganizing existing indicators—most of the time, assuming national systems of member states explicitly or implicitly as units of analysis. He warned that "important methodological difficulties await anyone interested in measuring intangibles like knowledge" (cf. Carter, 1996).

More recently, in the context of the preparation of FuturICT as a Flagship proposal to the European Commission, Helbing & Balieti (2011) took a more action-oriented approach to the problem of institutional barriers to the generation and diffusion of knowledge with a proposal to develop an "innovation accelerator" using bibliometric means. The scientometric method, in their opinion, has hitherto been too retrospective, whereas new techniques such as data mining, complexity studies, and artificial intelligence enable us to overcome barriers in the institutional information domain and lags in the system.

---

[1] The tenth edition of the *Science, Technology, and Technology Scoreboard 2011,* entitled "Innovation and Growth in Knowledge Economies," is available at http://www.oecd.org/document/10/0,3746,en_2649_33703_39493962_1_1_1_1,00.html.
[2] The statistics portal "Science, Technology, and Patents" of the OECD can be found at http://www.oecd.org/topicstatsportal/0,3398,en_2825_497105_1_1_1_1_1,00.html#500742.

For example, delays in publication processes can according to these authors be prevented by using and institutionalizing preprint servers (with quality control); and open access and open innovation models can stimulate the economy by providing more information in the public domain that can be used for innovation processes (Harnad, 2001). Although feedback loops are acknowledged by these authors (Kline & Rosenberg, 1986; cf. Fagerberg *et al.*, 2005), the main message is based on a linear, technology-push model of the innovation process: by enriching and speeding the information flow, innovation barriers can be washed away. Unlike economic control (e.g., in large corporations), information is considered as freely available.

Agarwal & Searls (2008, 2009) added to this information-driven and supply-side perspective on innovations, the option to data-mine the literature from the demand side using, for example, "Diseases" as need-articulation in the *Index Medicus*. Using the Medical Subject Headings (MeSH) in this index, one would be able to search the literature with the purpose of "literature-related discovery and innovation" (Swanson, 1990; Swanson & Smalheiser, 1999; cf. Kostoff, in press). Is it possible to retrieve relations between relevant literatures hitherto weakly connected and to exploit the strengths of these weak links (Granovetter, 1973) for innovation policies and R&D management? Can path-dependencies thus be generated and tunnels constructed under the divisions ("separatrices") among the different basins of attraction in science, technology, and innovation?

The strength-of-weak-links hypothesis is based on a structuralist perspective (Burt, 1992). The relations in the network span an architecture in a multi-dimensional space. This space can be mapped using, for example, techniques of multidimensional scaling (MDS). Within this space, however, specific points can be close to one another without necessarily being related in terms of network links (Leydesdorff & Rafols, 2012). For example, patent databases and scientific databases may develop with different rationales along their own axes with weak interactions between them. However, the positioning of the results in the common space of a single representation may enable us to specify how these different domains operate selectively upon each other, using measures such as structural holes (Burt, 1992) and/or betweenness centrality (Freeman, 1978/1979; cf. Leydesdorff, 2007). When is science relevant for technology, and when is this relation relevant for innovation? Note that the reverse arrow is also important given that research technologies can be considered as carriers of innovative trajectories both in science and the economy (Shinn, 2005).

**The scientometric perspective**
For the study of knowledge-based innovations, one needs to be able to move from representations of contexts of discovery to contexts of application, and *vice versa* (Gibbons *et al.*, 1994), in order to map path-dependencies, yet without losing control of how the interacting systems are further developed, both recursively and in relation to one another. Thus, we return to the problem of the different institutional contexts in which databases are maintained, classified, and made accessible to users as audiences with different knowledge interests, and therefore perspectives on the data.

The institutional incentives for accessing the Medline database, for example, are different for a medical practitioner confronted with the health problems of patients and for a laboratory scientist searching for references to support his/her knowledge claim. *Mutatis mutandis*, the same problem can be expected to occur in university-industry relations when one wishes to transfer knowledge from an academic to an industrial setting or, vice versa, translate demand articulation from industry to academia and into research programming. These different contexts can be expected to operate as selection environments asymmetrically upon each other during the process of innovation.

During the last decade, the various databases relevant to the innovation process have been investigated separately to a considerable extent. Much progress has been made in the mapping of science (Klavans & Boyack, 2009), and more recently overlay techniques have been developed that enable users to position document sets on maps, such as in overlays to Google Maps (e.g., Leydesdorff & Persson, 2010; Bornmann & Leydesdorff, 2011). The Derwent Innovation Index (DII) allows us to study patents and publications as well as citations among them in a single framework. The so-called "Non-literature patent references" (NLPR)—that is, references to literature other than patents—have been exploited in empirical studies (Glänzel & Meyer, 2003; Grupp, 1996; Narin & Noma, 1985; Narin & Olivastro, 1992; cf. Boyack & Klavans, 2008). However, the classifications and codifications in patent databases are very different from those in the scientific literature. Citations, for example, may mean something different in patents or scholarly literature because of the orientation towards legal protection as against reputation building in scholarly writing.

In an attempt to relate Google Maps of patents to scholarly literature, Leydesdorff & Bornmann (in press; cf. Bornmann & Leydesdorff, 2011) found the institutional address information a bottleneck in the relevant databases. The Derwent Innovation Index does not contain full address information; addresses of assignees in patent applications to the USPTO are often incomplete—but the addresses of inventors and the addresses in granted patents are complete and can be mapped (see at http://www.leydesdorff.net/patentmaps for an interactive tool)—and addresses in the bibliometric databases such as the Web-of-Science (WoS) and Scopus were found to be reliable to varying extents (Bornmann *et al.*, 2011). The address information in the Medline database is often confined to the corresponding author (Leydesdorff *et al.*, in preparation), but this selection is not systematic.

In addition to the geographic baseline map of Google Maps, scientometricians have mapped the different databases in terms of classifications or other socio-cognitively relevant aggregations such as journals or groups of journals representing specialties (Small & Garfield, 1985). The mapping of journals in terms of aggregated citation relations has a long tradition in scientometrics (e.g., Doreian & Farraro, 1985; Leydesdorff, 1986; Tijssen *et al.*, 1987). With the advent of enhanced visualization techniques, global maps of science could also be envisaged (Boyack *et al.*, 2005; de Moya-Anegón *et al.*, 2004; Leydesdorff & Rafols, 2009).

Rosvall & Bergstrom (2008) developed software that enables users to enter data online and draw maps from it (at http://www.mapequation.org/mapgenerator/index.html). This is a generic tool.[3] Rafols *et al.* (2010) developed dedicated software (available at http://www.leydesdorff.net/overlaytoolkit) that enables users to position one's sample in terms of the 220+ Subject Categories provided as representations of scientific specialties by Thomson Reuters, the current owner of the *Science Citation Index*. Similar baseline maps can be developed using citation patterns among patent classifications as indicators of intellectual organization (Newman *et al.*, 2011; Schoen *et al.*, 2011).

---

[3] See also http://www.leydesdorff.net/gmaps for using Pajek for geographic mapping of address information in terms of Google Maps.

**Interacting perspectives on the PubMed/Medline database**

Following Agarwal & Searls' (2009) suggestion, the perspectives of demand ("Diseases") and supply ("Drugs and Chemicals") are available as classifications to the PubMed/Medline database of the US National Institute of Health (NIH). This approach allows us to develop *different* baseline maps based on the *same* data, and their interaction in co-classifications. Additionally, a third branch of the index entitled "Analytical, Diagnostic and Therapeutic Techniques and Equipment" can be considered relevant to the process of medical innovations. However, one's mental map can be overburdened when change in three possible visualizations has to be related dynamically (Leydesdorff & Schank, 2008). Yet, animation techniques allow us to show the positions of clusters moving and relating in multivariate spaces. Recently developed visualization techniques allow such visualizations and animations, in principle, to be made interactive and web-based.

In a recent study, Leydesdorff *et al*. (in preparation) first show—using factor analysis on the basis of the complete document set 2010 of the PubMed database—that the three index branches relevant to the process of medical innovation can be considered as virtually independent of one another. In other words, one can obtain a visualization (comparable to Hofstadter's (1979) *Gödel-Escher-Bach* triplet) in which the projections are orthogonal. Using multidimensional scaling (MDS), however, one can lay out a map with the three domains as continents and visualize the MeSH terms in samples as overlays with corresponding colors. In Figure 1, for example, each dot represents one of the 822 second level MeSH terms of "Diseases" (red), "Drugs and Chemicals" (green), and "Analytical Diagnostic and Therapeutic Techniques and Equipment" (blue).

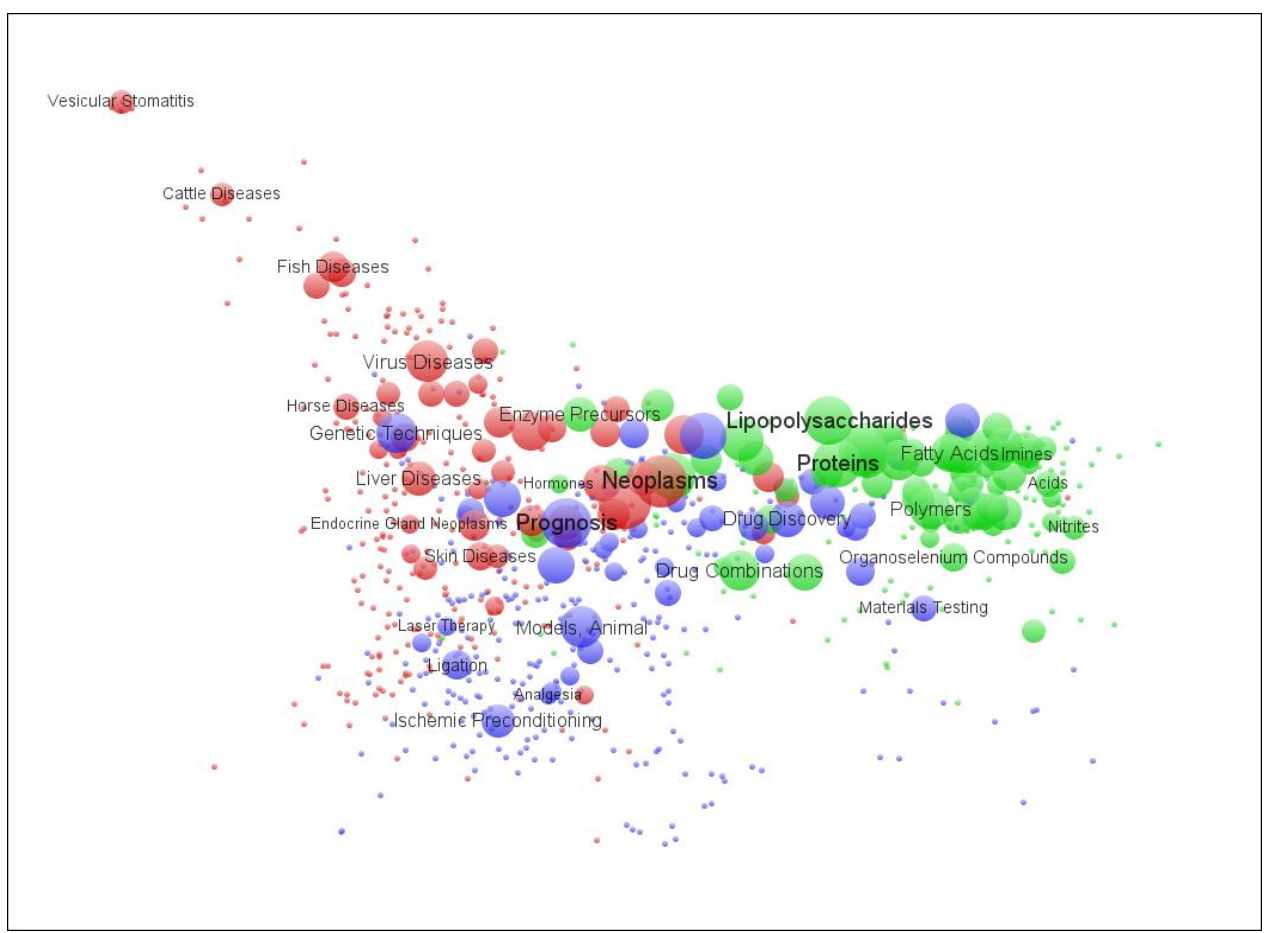

**Figure 1**: Overlay of 207 (among 822) MeSH terms on the base map of PubMed for 9,816 documents relevant to "RNA interference" in 2010; red represents "Diseases," green "Drugs and Chemicals," and blue "Techniques and Equipment". VOSViewer is used for the visualization. The size of the nodes is proportionate to the 2-logarithm of the number of documents (plus one).[4]

Specifically, Figure 1 projects the spread of "RNA interference" (RNAi) in the semantic domain of these second-level MeSH terms as classifiers of the 9,816 retrieved articles published in 2010.[5] By visualizing this overlay for different years, one can animate the development and show the spread of categories between 1998 and 2010 (available at http://www.leydesdorff.net/pubmed/rnai_vos.pps). The animation shows that the development after the scientific discovery in 1998 (Fire *et al.*, 1998) emerged first in the green domain of "Drugs and Chemicals," and then became increasingly relevant to all categories. In the period 2005-2007, for example, among the green-colored nodes the focus on "Prognosis" overshadowed the one on "Genetic Techniques".

As noted, we have discussed this specific development in a separate study (cf. Leydesdorff *et al.*, in preparation), but in the context of the argument here, an innovation can be represented as a trace in the database. In the case of RNAi, the trace begins at deeper levels of the index representing the R&D process where specialization prevails. The lower-level index terms can be integrated into this animation by collapsing the finer-grained categories at the second level. (For instructions on how to generate these maps, see http://www.leydesdorff.net/pubmed.)

During the innovation trajectory many traces will not survive, will fail to spread to other areas of the database, or will be "locked-in" (Arthur, 1989). Traces may also originate from techniques and equipment, or from new diseases shaping a research agenda (e.g., AIDS in the 1980s or SARS more recently; Leydesdorff *et al.*, 1994). Productive systems can be expected to generate variations continuously. The research question is, in our opinion, the specification of selection mechanisms, and how mutual selections may shape trajectories in co-evolutions.

**The integration of various databases**
In a paper entitled "The geography of science: disciplinary and national mappings," Small & Garfield (1985) noted that at least two baselines are possible for the mapping of science: a geographic one—which we operationalized above in terms of overlays to Google Maps—and the one of intellectual organization (cf. Whitley, 1984) that can be operationalized in the case of science, for example, in terms of journal-journal citations or co-citation maps (e.g., Small *et al.*, 1985). In recent years, important steps have been taken to integrate the mapping in different databases in terms of overlays to Google Maps; but mappings in terms of intellectual organizers cannot help being contextually bound by the markets/audiences of the databases under study. In other words, differentiation is found among the databases and their representations (Akera, 2007).

---

[4] The size of the nodes is proportional with the $\log_2(n+1)$ in order to prevent single occurrences ($n = 1$) from disappearing (because the $\log(1) = 0$).
[5] We used the following search string for the retrieval: "((((siRNA[Title/Abstract]) OR RNAi[Title/Abstract]) OR interference RNA[Title/Abstract]) OR RNA interference[Title/Abstract]) OR miRNA[Title/Abstract]) OR micro RNA[Title/Abstract]) OR interfering RNA[Title/Abstract])".

Would larger and more integrated datasets enable us to move back and forth through the data and thus map pathways and path-dependencies? Using RNAi as a marker in different settings, we traced this development in terms of publications and patents in the *Science Citation Index* and the USPTO database, respectively. In the scholarly context of publications—and comparing with nanocrystalline solar cells (NCSC) as a science-based development along the same time horizon—Leydesdorff & Rafols (2011) found first a change from a dynamics of preferential attachment to the inventors (1998-2002) to a next stage (2003-2010) in which attachment was preferential to emerging centers of excellence in metropolitan areas such as London, Boston, and Seoul.

In a study of transitions in innovation networks, Gay (2010) suggested to analyze these transitions in terms of the metaphor of Mark I and II in Schumpeter's models of entrepreneurial innovation. During the phase of Mark I, the entrepreneur leads the "creative destruction" of the old configurations, while in the period of Mark II "creative accumulation" by oligopolists—in this case, centers of excellence—can be expected to prevail (Soete & ter Weel, 1999). Unlike the Mode-1/Mode-2 distinction (Gibbons *et al*., 1994), these evolutionary dynamics are not generalized to the level of society, but technology and innovation-specific. Leydesdorff & Rafols (2011) suggest that in the case of the NCSC no transition to a diffusion dynamics across disciplinary boundaries was evident in the first decade of the 2000s.

When studying the same technology (RNAi) in terms of patents, using the USPTO database for the mapping of inventor addresses, the main centers of activity were unexpectedly found to be concentrated around Boulder and Denver, Colorado (Leydesdorff & Bornmann, in press; http://www.leydesdorff.net/patentmaps/sirna.htm). Whereas metropolitan centers (Boston-Cambridge, Houston-Austen-Dallas) are visible on the map, high-quality patenting (with citation rates in the top 10%) is concentrated in Colorado. In this study, the specific technology of RNAi was compared with nanotechnology. For the latter technology, U.S. patenting is concentrated above expectation in Silicon Valley. As a third comparison, the Netherlands was studied as a national system of innovations. In this case, major highways (between Amsterdam-Utrecht-Eindhoven and Amsterdam-The Hague-Rotterdam) were found to be axes of activity in (U.S.) patenting, whereas an expected cluster around the Agricultural University in Wageningen could not be retrieved (Porter, 2001: 43).

In other words, our results suggest that very different dynamics are at work which can be appreciated as the effects of different selection mechanisms (that have theoretically to be specified as hypotheses). Whereas scholarly papers can be expected to compete for attention and therefore citation, patents do not normally compete for citations: inventions are considered as non-rival in nature (Arrow, 1962; Romer, 1990). However, corporations compete in terms of patent portfolios.

Aggregation of patents showed the oligopolistic dominance of *Dharmacon RNAi Technologies* (Lafayette, CO) in patenting this new technology in the USA. Lundin (2011) studied RNAi in terms of both granted patents and patent applications, and in more databases than the USPTO (such as the database of the European Patent Office (EPO)). He noted stagnation in drug development because the problem of drug delivery *in vivo* is as yet insufficiently resolved. Patenting therefore has shifted to using RNAi technology as a reagent in other processes. Unlike the other firms in this market, which are drug-developing corporations (such as Merck), Dharmacon is a reagent supplier.

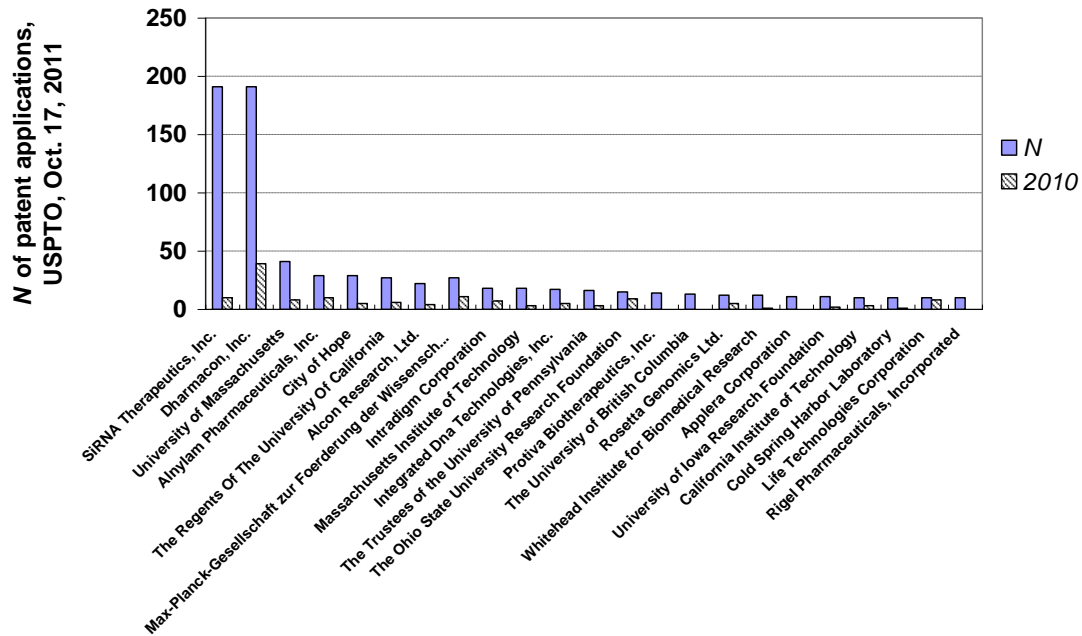

**Figure 2**: Assignees with ten or more applications among 2,343 patent applications with the USPTO since 2001; based on the database at http://appft.uspto.gov/netahtml/PTO/search-adv.html. Source: Figure 5 in Leydesdorff & Bornmann, in press.)

   Patent applications can be considered as closer to the research and invention process, because the granting of patents can sometimes take many years. Figure 2 shows that from the long-term perspective of patent applications, that is, including all years, siRNA Therapeutics owns as many patents as Dharmacon, and other players such as the University of Massachussetts (home of one of the co-inventors) and Alnylam Pharmaceuticals (the original spin-off company) also have substantive portfolios. However, a specific focus on 2010 shows the shake-out of the other companies by Dharmacon, also in terms of patent *applications*.

   In other words, the new technologies can move along trajectories in co-evolutions among all three relevant dimensions of the Triple Helix (geography, markets, and knowledge) and with potentially different dynamics. The globalization of the research front first required an uncoupling from the originators in the R&D process and a transition from Mode-1 to Mode-2 research in order to make the technique mutable (Latour, 1987, at pp. 226f). From this perspective, "Mode-1" and "Mode-2" are no longer considered as general systems characteristics of society and policy making, but as stages in the life-cycles of technological transformations. An analog of Schumpeter Mark I and Mark II within the domain of organized knowledge production and control can thus be specified.

   Universities are poorly equipped for patenting (Leydesdorff & Meyer, 2010). Some of the original patents may profitably be held by academia. In the case of RNAi, for example, two original US-patents ("Tuschl-I" and "Tuschl-II") were co-patented by MIT and the Max Planck Society in Germany (MIT Technology Licensing Office, 2006), but a company was founded as a spin-off to further develop the technology. The competition thereafter shifted along a commercial trajectory. In summary, whereas one can expect synergies to be constructed (Cooke & Leydesdorff, 2006), the consequent system "self-organizes" in terms of relevant selection environments, while leaving behind institutional footprints. Three dimensions are important: the economic, political/geographical, and socio-cognitive potentials for change. Both local integrations and global pressures for differentiation can continuously be expected.

**Towards an "Innovation Opportunity Explorer"**
Perhaps more modestly than an "innovation accelerator" (Helbing & Balieti, 2011), the scientometric perspective can guide us towards an "innovation opportunity explorer" by integrating heterogeneous datasets so that multiple perspectives can be related and pathways be suggested to policy makers and R&D management. However, one should remain aware that feasibility in a representation is different from realization. A focus presumes that certain other contexts will be considered as relatively stable. However, the abstraction may guide us in increasing our strategic awareness of new opportunities.

An innovation can to this end be conceptualized as the *trajectory* of an idea or concept within science as an intellectual and social organization as well as within the domain of legal encoding (patents) and marketing (industry). Ideas diffuse by way of carriers, e.g., papers, patents, and products, linking them to other ideas and to social actors: persons (authors, inventors, assignees) and organisations (scientific institutes, firms). Ideas evolve both substantively due to changing links with other ideas and organizationally during the trajectory. Evolutionary economics and technology studies offer (neo-Schumpeterian) models and hypotheses for the dynamics of both the substantive and organizational evolution, ICT offers tools for tracing trajectories in large datasets, while complexity science can clarify the system-level consequences of trajectory dynamics.

Because they are based on interactions among recursive selection mechanisms, trajectories of innovative ideas can be expected to consist of relatively fixed sets of steps or phases. Variation (knowledge claims, noise) is continuously filtered out. Our results (Leydesdorff & Rafols, 2011) suggest, for example, that medical innovations can begin as new ideas and empirical results (discoveries) within a single or a few isolated scientific groups, which, in a next phase, spread to many scientists (as in a chain reaction; Rogers, 1962). The scientists tend increasingly to connect to a few leading global institutes (oligopolistic centralization, creative accumulation). In this or a next stage, basic research can be complemented with interdisciplinary and translation research, e.g., in clinical trials, and patents begin to be registered.

This is a single example of a pattern likely to occur in the institutional trajectories of successful innovations; in a research programme such as FuturICT one could attempt to delineate more such patterns. Socio-cognitive patterns can then serve as signatures of innovation trajectories and their starting parts may allow for the identification of emerging and developing innovation trajectories. The latter can be prime candidates for targeted support when considered as early warning indicators.

More formally, and in terms of social network analysis, trajectories of innovative ideas can be conceptualized as temporally directed networks; network nodes are the carriers of the ideas while arcs represent the time-stamped diffusion of the ideas among carriers, for example, citation relations among papers or patents (Hummon & Doreian, 1989), and cooperation relations among persons or organizations. Bibliometric and semantic analysis can extract the networks from large databases of publications, patents, and so on. Efficient algorithms for the detection of signatures as small subnetworks in large sparse networks are available, e.g., implemented in Pajek software (De Nooy *et al.*, 2005). These algorithms can be developed to handle the temporal dimension of signatures and the multi-relational character (cooperation, citation, co-citation, co-affiliation, concordance, alliances, and so on) of the networks (Leydesdorff, 2010b). This new type of fragment detection can be considered as the social-scientific counterpart of sequencing techniques in the sciences (Abbott, 1995).

Whereas the dynamics of innovation trajectories can thus be reconstructed at the micro level of ideas, persons, and institutes, the neo-Schumpeterian hypotheses specify conditions that foster or impede the development of an innovation trajectory. Examples include the institutional inertia hypothesis (Agarwal & Searls, 2009), the preferential attachment (winner-take-all) hypothesis (Barabási & Albert, 1999; Price, 1976), and "lock-in" and hyperstabilization along a trajectory (Arthur, 1989) versus meta-stabilization and globalization as a regime (Dolfsma & Leydesdorff, 2010). Statistical network models enable us to assess the strength of these effects on the development of trajectories at the micro level (De Nooy, 2011; Kolaczyk, 2009); substantive effects are potential candidates for policy-based interventions.

The step towards action and intervention requires further reflection because the retention of wealth from knowledge (or knowledge from wealth) presumes the specification of an institutional and/or geographic system of reference with a dynamics of its own. Note that the innovation tract is heavily institutionalized. Furthermore, local nonlinear dynamics can be expected to give rise to complex systems and possibly unexpected and unwanted outcomes at the systems level. Targeted local interventions, e.g., improving the conditions for a research group, may not be effective if progress depends on the network context.

Complexity science, multi-agent and stochastic simulation models are needed to evaluate the performance of innovation trajectories at the systems level and their susceptibility or resilience to changing conditions, that is, changes in parameter values at the micro level. The statistical estimates of the trajectory parameters can be used to calibrate the simulation models. Thus, the ambition of FuturICT to combine information-theoretical models, statistical network models, and simulation models of complex systems results in a multi-level complex model rooted in the social sciences. One combines behavioural hypotheses at the micro level of agents with recursive self-organization of knowledge at the systems level. The scientometric perspective contributes to the over-arching problems of ecological and social mechanisms in complex phenomena by focusing on the process of nonlinear innovations in knowledge-based economies.

**Acknowledgements**

We thank Ismael Rafols for comments on a previous draft, and acknowledge support by the ESRC project 'Mapping the Dynamics of Emergent Technologies' (RES-360-25-0076).